\documentclass[twocolumn, twocolappendix]{aastex631}

\usepackage{acronym}
\usepackage{xspace}
\usepackage{amsmath}
\usepackage{hhline}
\usepackage{appendix}
\usepackage{listings}

\lstdefinestyle{bash}{
  language=bash,
  basicstyle=\ttfamily\scriptsize,
  numbers=left,
  numberstyle=\tiny\color{gray},
  stepnumber=1,
  numbersep=5pt,
  backgroundcolor=\color{white},
  showspaces=false,
  showstringspaces=false,
  showtabs=false,
  frame=none,
  rulecolor=\color{black},
  tabsize=4,
  captionpos=b,
  breaklines=true,
  breakatwhitespace=false,
  commentstyle=\color{blue},
  morekeywords={ls, cd, cp, mv, rm, mkdir, rmdir, touch, chmod, chown, sudo, apt, yum, echo, cat},
}

\acrodef{bbh}[BBH]{binary black hole}
\acrodef{imbh}[IMBH]{intermediate-mass black hole}
\acrodef{bns}[BNS]{binary neutron star}
\acrodef{FAR}[FAR]{false alarm rate}
\acrodef{bf}[BF]{Bayes' factor}
\acrodef{cbc}[CBC]{compact binary coalescence}
\acrodef{ce}[CE]{Cosmic Explorer}
\acrodef{SNe}[SNe]{Supernova}
\acrodef{da}[DA]{data analysis}
\acrodef{et}[ET]{Einstein Telescope}
\acrodef{eob}[EOB]{Effective-One-Body}
\acrodef{fd}[FD]{frequency domain}
\acrodef{gw}[GW]{gravitational-wave}
\acrodef{gr}[GR]{general relativity}
\acrodef{hm}[HM]{Higher mode}
\acrodef{ifo}[IFO]{interferometer}
\acrodef{imr}[IMR]{inspiral-merger-ringdown}
\acrodef{im}[IM]{inspiral-to-merger}
\acrodef{kagra}[KAGRA]{Kamioka Gravitational Wave Detector}
\acrodef{ligo}[LIGO]{Laser Interferometer Gravitational-Wave Observatory}
\acrodef{lso}[LSO]{Last Stable Orbit}
\acrodef{lvc}[LVC]{LIGO-Virgo Collaboration}
\acrodef{lvk}[LVK]{LIGO-Virgo-Kagra Collaboration}
\acrodef{lo}[LO]{leading order}
\acrodef{ns}[NS]{neutron star}
\acrodef{bh}[BH]{Black hole}
\acrodef{nr}[NR]{numerical relativity}
\acrodef{pn}[PN]{post-Newtonian}
\acrodef{pe}[PE]{parameter estimation}
\acrodef{psd}[PSD]{power spectral density}
\acrodef{KN}[KN]{kilonova}
\acrodef{2g}[2G]{second-generation}
\acrodef{xg}[XG]{next-generation}
\acrodef{em}[EM]{electromagnetic}
\acrodef{bhns}[BHNS]{black-hole neutron star binary}
\acrodef{asd}[ASD]{amplitude spectral density}
\acrodef{agn}[AGN]{active galactic nuclei}

\acrodefplural{KN}[KNe]{kilonovae}
\acrodef{qc}[QC]{quasi-circular}
\acrodef{snr}[SNR]{signal-to-noise ratio}
\acrodef{SNR}[SNR]{signal-to-noise ratio}
\acrodef{ng}[NG]{Next Generation}
\acrodef{eos}[EoS]{Equation of State}

\newcommand{\pycbc}{\textsc{PyCBC}\xspace}

\newcommand{\nrsur}{\texttt{NRSur7dq4}\xspace}

\newcommand{\bilby}{\texttt{Bilby}\xspace}
\newcommand{\dynesty}{\texttt{dynesty}\xspace}
\newcommand{\gwforge}{\texttt{gwforge}\xspace}
\newcommand{\git}{\texttt{git}\xspace}

\newcommand{\population}{\texttt{population}\xspace}
\newcommand{\ifo}{\texttt{ifo}\xspace}
\newcommand{\inject}{\texttt{inject}\xspace}

\newcommand{\Gpcyr}{\ensuremath{\mathrm{/Gpc^3/yr}}}

\begin{document}

\title{\gwforge: A user-friendly package to generate gravitational-wave mock data}

\correspondingauthor{kbc595@psu.edu}
%\email{greg.schwarz@aas.org, gus.muench@aas.org}

\author[0000-0003-4750-5551]{Koustav Chandra}
\affiliation{Institute for Gravitation \& the Cosmos and Department of Physics, The Pennsylvania State University, University Park PA 16802, USA}



\begin{abstract}

The \acl{xg} \acl{gw} detectors, with their improved sensitivity and wider frequency bandwidth, will be capable of observing almost every \acl{cbc} signal from epochs before the first stars began to form, increasing the number of detectable binaries to hundreds of thousands annually. This will enable us to observe compact objects through cosmic time, probe extreme matter phenomena, do precision cosmology, study gravity in strong field dynamical regimes and potentially allow observation of fundamental physics beyond the standard model. However, the richer data sets produced by these detectors will pose new computational, physical and astrophysical challenges, necessitating the development of novel algorithms and data analysis strategies. To aid in these efforts, this paper introduces \gwforge, a user-friendly, lightweight \textsc{Python} package, to generate mock data for \acl{xg} detectors. We demonstrate the package's capabilities through data simulation examples and highlight a few potential applications: performance loss due to foreground noise, bright-siren cosmology and impact of waveform systematics on binary \acl{pe}.
% \RG{Also: this abstract contains a lot of ``indtroduction'' about next-gen, but not much discussion of the actual results. This might be fine though -- what is the target journal?}
\end{abstract}

%%%%%%%%%%%%%%%%%%%%%%%%%%%%%%%%%%%%%%%%%%%%%%%%%%
\section{Introduction} \label{sec:intro}
%%%%%%%%%%%%%%%%%%%%%%%%%%%%%%%%%%%%%%%%%%%%%%%%%%

% \RG{Overall comment: great job! cool paper, there are only a few things that I think may require a bit of further clarification/explanation. 
% \begin{itemize}
    % \item Personally, I would put some more emphasis on the fact that with XG detectors you have inject a ton of ``foreground'' signals to generate a realistic set of data.
    % \item As a direct consequence of the comment above, I miiiiight suggest to add at the end of Sec. III one figure showing the time domain data with noise + BBH + BNS + BHNS. The BBHs should (?) be visible, at least the mergers.
    % \item You can then say that you used the data generated/shown at the end of Sec.III to do all of the analyses in sec IV (if this is true, ofc)
% \end{itemize} 
% I left some other comments here and there, but not much.}

The \acf{xg} ground-based \acf{gw} detectors, such as the \ac{ce} and the \ac{et}, promise to significantly improve both the detection horizon and the frequency bandwidth of observable \ac{cbc} signals~\citep{Punturo:2010zz, Reitze:2019iox, Maggiore:2019uih, Evans:2021gyd, Borhanian:2022czq, Branchesi:2023mws, Evans:2023euw, Gupta:2023lga}. This expanded detection horizon will allow us to observe \acf{cbc} signals from greater distances, thereby increasing the number of detectable binaries by several orders of magnitude. Additionally, the increased bandwidth of these detectors could allow for the observation of a \ac{cbc} signal over durations spanning minutes to even hours~\citep{Regimbau:2012ir, Evans:2021gyd, Gupta:2023lga}. 

Given this, the \ac{xg} \ac{gw} detectors will have far-reaching scientific potential. However, the ``richer'' \ac{gw} detector data will present computational, physical and astrophysical challenges. Existing \ac{gw} analysis tools, waveform models and detector calibration techniques will be insufficient in \ac{xg} era~\citep{Evans:2021gyd, Pizzati:2021apa, Hu:2022bji}. Also, the sheer volume of observed signals will necessitate the development of novel algorithms for detecting the signals and extracting the science out of them. It is, therefore, imperative to quantify the limitations of current tools when employed in \ac{xg} data analysis. 
% \VC{Might be helpful to broadly mention what the referenced papers conclude about why current tools will be ``insufficient.''} 

To support this effort and ensure that \ac{xg} detectors will achieve their scientific potential, it is crucial to generate playground datasets to allow the \ac{gw} community to develop and tune their algorithms. Such an effort will drive the development of new data analysis strategies, such as faster search and \ac{pe} techniques, improved waveform modelling, dealing with non-stationary noise and methods to identify subtle physical effects, such as finite size effects in \ac{ns}. 

This paper introduces \gwforge, a user-friendly \textsc{Python} package to simulate the mock data for the proposed \ac{xg} detectors based on our understanding of our universe. 
%It provides primary investigation tools \IG{elaborate here?} with straightforward syntax, enabling simple analyses of the simulated datasets. 
One primary reason for developing \gwforge is the anticipated abundance of signals in \ac{xg} data, necessitating systematic data generation that accounts for realistic population models and the complexities of overlapping signals. This code is built on existing \ac{gw} packages, such as \pycbc, \bilby, \texttt{gwpy} and \texttt{gwpopulation}.
% \VC{would improve clarity to split this sentence like ``… \ac{xg} data. This necessitates systematic …’’}
% \IG{The last sentence cannot be the final sentence to this para. In general, this para should be bigger, IMO giving a more detailed overview.}

The subsequent sections of this paper are structured to highlight the user-friendly nature of the code. Sec~\ref{sec:overview} provides an overview of the code, including installation instructions in Sec.~\ref{sec:install}. Sec.~\ref{sec:day} provides the necessary theoretical background and an example to generate a day's worth of coloured Gaussian \ac{xg} network data with \ac{cbc} signals. Sec.~\ref{sec:application} is devoted to a few studies one can do using the prepared dataset in the previous section, and we conclude in Sec.~\ref{sec:conclusion}.

% . Finally,   has the closing remarks.
% \RG{Wdym by ``status of the code"?}
% \RG{Also: I think you don't mention one of the (maybe most important?) reasons we need  with gwforge: the fact that there will be a TON of overlapping signals, which means that current ways of generating mock GW data (eg with bilby) are not sufficient. Hence, there is need of a dedicated package, to systematically do this taking into account realistic population models etc etc}
    
%%%%%%%%%%%%%%%%%%%%%%%%%%%%%%%%%%%%%%%%%%%%%%%%%%
\section{Code Overview}\label{sec:overview}
%%%%%%%%%%%%%%%%%%%%%%%%%%%%%%%%%%%%%%%%%%%%%%%%%%

%%%%%%%%%%%%%%%%%%%%%%%%%%%%%%%%%%%%%%%%%%%%%%%%%%
\subsection{Installation}\label{sec:install}
%%%%%%%%%%%%%%%%%%%%%%%%%%%%%%%%%%%%%%%%%%%%%%%%%%
The source code is available on the \git repository \citep{gwforge-git}, which includes instructions for contributing to code development. Comprehensive documentation regarding installation, functionality, and user syntax is provided on the project’s website \citep{gwforge-doc}. Example setup scripts discussed in this work are available in the \git repository. %\RG{Might be worth to make the code pip-installable or conda-forge-able... for the lazy user, eg myself :)} \KC{I need help with that. :) I believe making it conda installable will be awesome}

%%%%%%%%%%%%%%%%%%%%%%%%%%%%%%%%%%%%%%%%%%%%%%%%%%
\subsection{Packages}
%%%%%%%%%%%%%%%%%%%%%%%%%%%%%%%%%%%%%%%%%%%%%%%%%%

At the top level, \gwforge comprises three main packages: \population, \ifo and \inject. The \population package provides functions essential to generate individual compact binary properties $\boldsymbol{\theta}$ based on hyper-parameters $\boldsymbol{\Lambda}$ that describe the population characteristics. These characteristics are defined by phenomenological population models or detailed physical simulations, such as population synthesis or N-body dynamical simulations generated using population synthesis codes such as \textsc{Compas}~\citep{Stevenson:2017tfq, Vigna-Gomez:2018dza}. \gwforge integrates all phenomenological population models from \texttt{gwpopulation} and includes additional distribution functions such as a \texttt{DoubleGaussian} and \texttt{LogNormal}. Users can also utilise outputs from population synthesis simulations to generate their own populations and define cosmological parameters, such as the Hubble constant $H_0$, the CMB temperature $T_\mathrm{CMB}$, etc., to specify their Universe.

The \ifo package provides core functionality for generating either zero-noise or coloured Gaussian noise given a noise \ac{psd}. It implements current and future ground-based \ac{gw} detectors, detailing their locations, orientations, and various noise PSDs of existing and future instruments. Users can also implement custom \ac{gw} detectors akin to \bilby~\citep{Ashton:2018jfp}. Additionally, this package allows users to introduce data gaps, enabling the simulation of detector downtimes. This is important, as it creates a more realistic scenario, reflecting the potential for missing signals or having signals that are only partially present. By accounting for detector downtimes and data gaps, one can better understand the impact of these factors on gravitational wave detection and analysis, thereby improving the robustness of signal detection and parameter estimation methodologies. 

% \IG{maybe say why this is important, like, this is a more realistic scenario, we might miss of signals, some signals may only be partially present, etc.}.

Finally, the \texttt{inject} module provides essential functionalities for \textit{injecting} or adding simulated \ac{gw} signals into the data. It supports all standard waveform approximants available via the \texttt{lalsimulation} package~\citep{lalsuite} and/or those made available via \pycbc's waveform plugin.

Future updates will expand this module to include the capability to add non-Gaussian transients, \ac{SNe} signals, and other \ac{gw} transients, thereby making the network output more closely ``mock'' the real data.

%%%%%%%%%%%%%%%%%%%%%%%%%%%%%%%%%%%%%%%%%%%%%%%%%%
\subsection{Executables}
%%%%%%%%%%%%%%%%%%%%%%%%%%%%%%%%%%%%%%%%%%%%%%%%%%

To facilitate usability, we developed three key executables using the aforementioned packages: \texttt{gwforge\_population}, \texttt{gwforge\_noise}, and \texttt{gwforge\_inject}. Each executable operates based on a configuration file tailored to its specific function. Additionally, the \texttt{gwforge\_workflow} executable integrates all three tools and utilises the high-throughput computing framework HTCondor. This integration significantly accelerates the data generation process and simplifies user interaction by managing all ancillary tasks automatically. 
%
% \RG{I would put the following sentence at the beginning of the following section}
The following section provides an explicit example of using these executables.

%%%%%%%%%%%%%%%%%%%%%%%%%%%%%%%%%%%%%%%%%%%%%%%%%%
\section{Simulating a day's worth of mock data}\label{sec:day}
%%%%%%%%%%%%%%%%%%%%%%%%%%%%%%%%%%%%%%%%%%%%%%%%%%

%%%%%%%%%%%%%%%%%%%%%%%%%%%%%%%%%%%%%%%%%%%%%%%%%%
\subsection{Background}\label{sec:background}
%%%%%%%%%%%%%%%%%%%%%%%%%%%%%%%%%%%%%%%%%%%%%%%%%%

The observed \ac{gw} strain \( s(t \mid \boldsymbol{\theta}, I) \) at a detector \( I \) can be expressed as:
\begin{equation}
    s(t \mid \boldsymbol{\theta}, I) = \frac{1}{D_L} \mathbb{R}\left[F^I h\right]
\end{equation}
where \(\boldsymbol{\theta}\) is a vector representing the source's properties,  \( D_L \) is the luminosity distance to the source, $\mathbb{R}$ denotes the real part, and
\[ 
F = F_+^I + iF_\times^I % \VC{Should this be F^I?}
\]
is the antenna response function that depends on the sky location \((\alpha, \delta)\) and the polarization angle \((\psi)\). The term \( h \) represents the complex binary waveform on a sphere. This waveform can be decomposed as a sum of spin-weighted spherical harmonic modes \( h_{\ell m} \), so that the waveform along any direction \((\iota, \varphi)\) in the binary's source frame is given by:
\begin{equation}
    h = \sum_{\ell \geq 2} \sum_{m=-\ell}^{\ell} {}^{-2}{Y}_{\ell,m }(\iota, \varphi) h_{\ell m}(t \mid \boldsymbol{\Xi})
\end{equation}
Here, \( {}^{-2}{Y}_{\ell,m }(\iota, \varphi) \) are the spin $-2$ weighted spherical harmonics, and \((\iota, \varphi)\) represent the binary's inclination and azimuth, respectively. The vector \(\boldsymbol{\Xi}\) denotes the binary's intrinsic parameters whose dimensionality depends on the binary type. For example, in quasi-spherical \acp{bbh}, \(\boldsymbol{\Xi}\) includes the two component masses \(m_i\) and the six spin degrees of freedom \(\boldsymbol{\chi}_i\). In contrast, for quasi-circular or non-precessing \acp{bns}, \(\boldsymbol{\Xi}\) represents the component masses \(m_i\),  the tidal parameters \(\Lambda_i\) and the aligned spin components \(\boldsymbol{\chi}_i \cdot \hat{L}\), with $\boldsymbol{L}$ being the binary's orbital angular momentum.

Since the source's total merger rate and redshift distribution determine the number of signals in a given simulated dataset, \gwforge begins by first estimating the average time interval \(\langle \Delta t \rangle\) between two adjacent signals of the same type given the source population and the cosmological parameters. It assumes that all \ac{cbc} systems are formed through isolated formation channels involving a common envelope phase. This allows it to derive the redshift distribution of \acp{cbc} in the source frame as follows~\citep{Zhu:2020ffa, Wu:2022pyg}:
\begin{equation}
    \begin{aligned}
\mathcal{R}(z) &\propto  \int_{\tau_{\text{min}}}^{\infty} \mathcal{R}_\ast [z_f(z, \tau)] P(\tau)d\tau \\  
&\propto \int_{z}^{\infty} \mathcal{R}_\ast (z_f) P[\tau(z, z_f)] \frac{dt(z_f)}{dz_f} \, dz_f~.
    \end{aligned}
\end{equation}
Here, \(\mathcal{R}_\ast\) is the cosmological star formation rate density, \(P(\tau)\) is the probability density function of the delay time, \(z_f\) is the redshift at which the stellar binary forms, \(z\) is the redshift at which the compact binary merges, \(\tau=t(z)-t(z_f)\) is the delay time and \(t(z)\) is the lookback time at redshift $z$. The proportionality constant is defined such that $\mathcal{R}(z=0) = \mathcal{R}_0$ where \(\mathcal{R}_0 \) is the local merger rate density of the \ac{cbc} source.

The lookback time,
\begin{equation}
    t(z) = \frac{1}{H_0} \int_{z}^{\infty} \frac{dz}{(1+z) \sqrt{\Omega_\Lambda + \Omega_m(1+z)^3}},
\end{equation}
depends on the Hubble constant, \(H_0\), the dark matter, \(\Omega_\Lambda \) and matter density, \(\Omega_m\). All these cosmological parameters, along with the star formation rate, are adjustable by the user, allowing them to simulate their own Universe. However, \gwforge currently only implements \(P(\tau) \propto 1/\tau \).

Given this, the merger rate per redshift bin in the detector frame is then given by~\citep{Regimbau:2012ir}:
\begin{equation}\label{eq:rate}
    \frac{dR}{dz} = \frac{\mathcal{R}}{1+z} \frac{dV(z)}{dz},
\end{equation}
where the comoving volume element is given as:

\begin{equation}
\frac{dV(z)}{dz} = \frac{c}{H_0} \frac{4 \pi D_L^2}{(1+z)^2 \sqrt{\Omega_\Lambda + \Omega_m(1+z)^3}}~.
\end{equation}\\

Here, \(c\) is the light speed, and \(D_L\) is the luminosity distance to the source defined as:
\begin{equation}
D_L =  \frac{c(1+z)}{H_0} \int_0^z \frac{dz'}{\sqrt{\Omega_\Lambda + \Omega_m(1+z')^3}}
\end{equation}
The $1/(1+z)$ factor in Eq.~\eqref{eq:rate} accounts for time dilation.

\gwforge obtains \(\langle \Delta t \rangle\) as follows:
\begin{equation}
    \langle \Delta t \rangle = 1 / \Big( \int^{z_\mathrm{max}}_0 \frac{dR}{dz} dz \Big)~.
\end{equation}
where $z_\mathrm{max}$ is the maximum redshift used in the simulation. Based on \(\langle \Delta t \rangle\) and the total analysis time, \(T_a\) of the network, \gwforge determines the expected number of signals, \(N = T_a / \langle \Delta t \rangle\), and correspondingly draws the binary parameters from the provided population models.

Below, we discuss an example to build a mock catalogue based on our current understanding of \ac{bbh} binary population distribution. 

%%%%%%%%%%%%%%%%%%%%%%%%%%%%%%%%%%%%%%%%%%%%%%%%%%
\subsection{Simulating the mock GW catalogue}
%%%%%%%%%%%%%%%%%%%%%%%%%%%%%%%%%%%%%%%%%%%%%%%%%%
To create a mock \ac{bbh} catalogue using \gwforge, one must first define a configuration file like the one provided below:
\begin{widetext}
    \begin{lstlisting}[style=bash]
[Redshift]
redshift-model = MadauDickinson
redshift-parameters = {`gamma': 2.7, `kappa': 5.6, `z_peak': 1.9}
local-merger-rate-density = 22
maximum-redshift = 30
; custom cosmology
cosmology = custom 
H0 = 67.7
Om0 = 0.31
Ode0 = 0.69
Tcmb0 = 2.735
; analysis start time
gps-start-time = 1893024018

[Mass]
mass-model = PowerLaw+Peak
mass-parameters = {`alpha':3.37, `beta': 0.76, `delta_m':5.23,  `mmin':4.89, `mmax':88.81, `lam':0.04, `mpp': 33.60, `sigpp':4.59}

[Spin]
spin-model = default
spin-parameters = {`mu_chi':0.26, `sigma_squared_chi':0.02, `sigma_t':0.87, `xi_spin':0.76}

[Extrinsic]
    \end{lstlisting}
\end{widetext}

Within the \texttt{[Redshift]} section, users must specify the redshift evolution model, local merger rate density in \Gpcyr, analysis start time (in GPS time), and the maximum redshift. Currently, \gwforge supports two different redshift evolution models: the power-law redshift evolution model and the other that convolves the Madau-Dickinson star-formation rate with an inverse time-delay model to determine the \ac{cbc}'s redshift evolution model~\citep{Fishbach:2018edt}. Users can define custom cosmological and redshift parameters; otherwise, default values from \citet{Planck:2018vyg} and Madau-Dickinson parameters are used~\citep{Madau:2014bja}.

Similarly, the \texttt{[Mass]} and \texttt{[Spin]} sections provide options for various phenomenological mass and spin distribution models, as detailed in \citet{gwforge-doc}. Besides predefined models, \gwforge can utilize outputs from population-synthesis codes, creating interpolated distribution functions from variable and associated probability arrays, enabling sampling from simulated distributions.

Similarly, the \texttt{[Mass]} and \texttt{[Spin]} sections offer a selection of phenomenological mass and spin distribution models, as documented in \citet{gwforge-doc}. 
% Besides predefined models, \gwforge can utilise outputs from population-synthesis codes, creating interpolated distribution functions from variable and associated probability arrays, enabling sampling from simulated distributions. \IG{Sounds like a repetition of the last paragraph.}
The \texttt{[Extrinsic]} section deals with the \textit{other} extrinsic parameters. Unless otherwise specified, \gwforge employs the following default distribution choices for these parameters (See Table~\ref{tab:true-values}) for definitions of the symbols):
\begin{table}[htb]
    \centering
\begin{tabular}{ccccc}
\hline \hline variable & unit & distribution & minimum & maximum \\
\hline$\alpha$ & rad. & uniform & 0 & $2 \pi$ \\
$\delta$ & rad. & $\cos$ & $-\pi / 2$ & $\pi / 2$ \\
$\iota$ & rad. & sin & 0 & $\pi$ \\
$\psi$ & rad. & uniform & 0 & $\pi$ \\
$\phi$ & rad. & uniform & 0 & $2 \pi$ \\
\hline \hline
\end{tabular}
\caption{The default distribution functions for the sky location parameters, $(\alpha, \delta)$, inclination, $\iota$, polarisation angle $\psi$ and azimuth $\phi$.}
    \label{tab:my_label}
\end{table}

% To enhance flexibility, \gwforge enables users to define their own distributions akin to those utilised in \bilby, thus facilitating tailored simulations according to specific preferences. 
% \IG{I feel that the point about user-defined distributions can be made once and for all, because right now it feels very repetitive.}

Finally, if the binary contains a \ac{ns} or a pair of \acp{ns}, \gwforge can set the tidal parameters $\Lambda_i$ based on an \ac{eos}. This necessitates a pre-shipped \texttt{ASCII} table, examples of which can be found at \citet{gwforge-git}. Additional configuration file examples are also provided at the same location.

Once the configuration file is defined, the next step is to execute:
\begin{widetext}
\footnotesize
\begin{lstlisting}[style=bash]
    gwforge_population --config-file bbh.ini --output-file bbh.h5
\end{lstlisting}
\end{widetext}
to create the mock \ac{bbh} catalogue and store as in an \texttt{hdf5} file.

%%%%%%%%%%%%%%%%%%%%%%%%%%%%%%%%%%%%%%%%%%%%%%%%%%
\subsection{Mock noise generation}
%%%%%%%%%%%%%%%%%%%%%%%%%%%%%%%%%%%%%%%%%%%%%%%%%%

After generating the mock \ac{gw} catalogue, the next step is to define the detector network and simulate their noise. This can be achieved by creating a configuration file, for example, \texttt{xg.ini}, as shown below:
\begin{widetext}
    \begin{lstlisting}[style=bash]
[IFOS]
detectors = [`CE20', `CE40', `ET']
sampling-frequency = 8192
noise = gaussian
    \end{lstlisting}
\end{widetext}
and executing:\\
\begin{widetext}
\begin{lstlisting}[style=bash]
    gwforge_noise --config-file xg.ini --output-directory output/data --gps-start-time 1895616018 --gps-end-time 1895656978
\end{lstlisting}
\end{widetext}
This initialises a three-detector network consisting of two L-shaped \ac{ce} observatories at two fiducial locations: \ac{ce}40 off the coast of Washington State and \ac{ce}20 off the coast of Texas, along with an underground triangular \ac{et} located in Sardinia, Italy~\citep{Evans:2023euw, Gupta:2023lga}. The simulated noise thus produced will be Gaussian coloured by the detector noise \ac{psd}, with data sampled at 8192 Hz over approximately one day. The data will be stored as \texttt{HDF5} files in the \texttt{output/data} directory, named in the format \texttt{{IFO}-{GPS-TIME}.h5}, with channel names \texttt{{IFO}:INJ}. To simulate zero-noise realisations, the user has to set the flag \texttt{noise = zero}.

\gwforge also includes the two LIGO detectors in the US~\citep{LIGOScientific:2014pky}, Virgo in PISA~\citep{VIRGO:2014yos}, LIGO-Aundha in India~\citep{LIGOIndia} and the KAGRA detector in Japan~\citep{Aso:2013eba}. It is also possible to define new detectors characterised by their geometry, location and frequency response. This feature is particularly useful for developing the science case for new proposals and optimising the design and placement of new detectors, such as placing the \ac{et} observatory in the Netherlands or building a pair of \ac{et} observatories at two different locations~\citep{Branchesi:2023mws}.
% \RG{Is it possible to change the location of pre-defined observatories?  Eg say I want my ET to be in the Netherlands, do I have to define a new ifo or is it possible to just change the location of the current ET?}
%%%%%%%%%%%%%%%%%%%%%%%%%%%%%%%%%%%%%%%%%%%%%%%%%%
\subsection{Mock data generation}
%%%%%%%%%%%%%%%%%%%%%%%%%%%%%%%%%%%%%%%%%%%%%%%%%%

\begin{figure*}
    \centering
    \includegraphics[width=\textwidth]{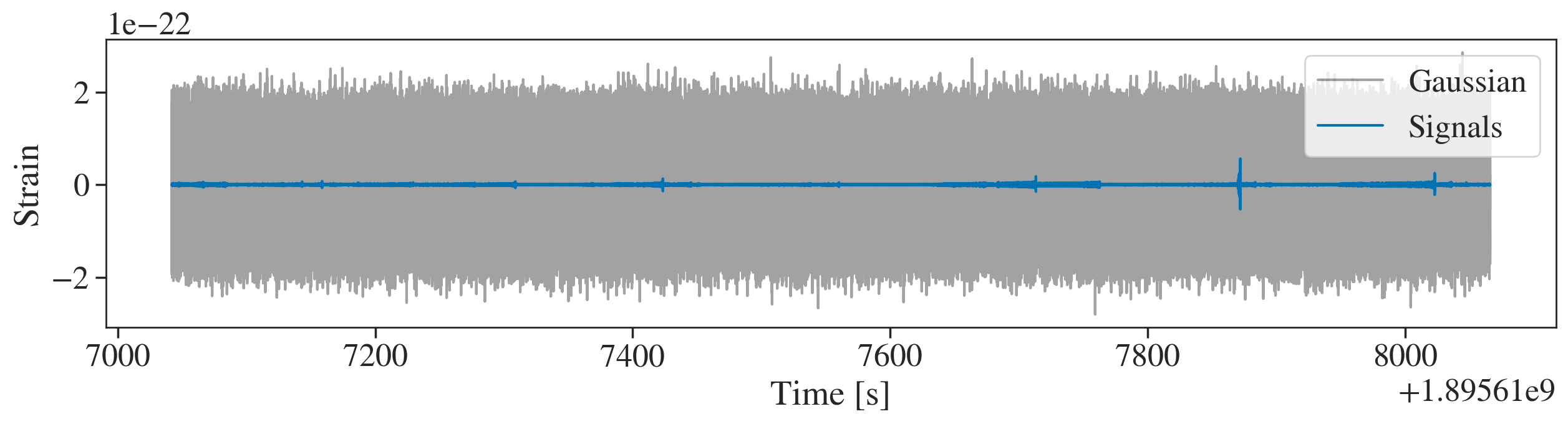}
    \caption{A representative section of data generated using \gwforge. The grey shows the Gaussian noise while the blue represents the sum of all \ac{gw} signals, namely from \ac{bns}, \ac{bhns} and \ac{bbh}.}
    \label{fig:data}
\end{figure*}

The final step involves injecting the simulated \ac{gw} signals into the network data. Similar to the previous steps, this requires a configuration file, for example, \texttt{inject-bbh.ini}, as shown below: 

\begin{widetext}
\begin{lstlisting}[style=bash]
[IFOS]
detectors = [`CE20', `CE40', `ET']
channel-dict = {`CE20':`CE20:INJ', `CE40':`CE40:INJ', `ET':`ET:INJ'}
sampling-frequency = 8192
minimum-frequency = 6

[Injections]
injection-file = bbh.h5
injection-type = bbh
waveform-approximant = IMRPhenomXO4a
\end{lstlisting}
\end{widetext}

As with the noise generation step, the \texttt{[IFOS]} section requires specification of the detector network and sampling frequency. Additionally, this section must include the minimum frequency cut-off for signal generation and the channel names for the frame files. If not specified, \gwforge assumes that the reference frequency for measuring the spin angles is the same as the minimum frequency used for signal generation.

The \texttt{[Injections]} section should contain the following keys: \texttt{injection-file}, which points to the file containing the mock \ac{bbh} catalogue (e.g., \texttt{bbh.h5}); \texttt{injection-type}, which specifies the type of injection being made; and \texttt{waveform-approximant}, which indicates the waveform model to use for signal generation, with \texttt{IMRPhenomXO4a} being an example~\citep{Thompson:2023ase}. Currently, \gwforge supports all waveform approximants available via the \texttt{lalsimulation} package~\citep{lalsuite}. To use waveforms beyond those available in \texttt{lalsimulation}, a new plugin package must be created to advertise the new waveform model to \pycbc for waveform generation~\citep{Usman:2015kfa}.

% \RG{So this means that you use pycbc's get-td-waveform to gen signals, is this right?}
% \KC{not quite  --- I give the user the choice --- they can either use PyCBC or Bilby or lalsimulation :D }

To inject the signal waveforms into the generated data over the specified GPS time interval, the user needs to execute the following command:
\begin{widetext}
\begin{lstlisting}[style=bash]
gwforge_inject --config-file forged-data/inject-bbh.ini --data-directory /output/data --gps-start-time 1895616018 --gps-end-time 1895656978        
\end{lstlisting}
\end{widetext}

This command injects the signal waveforms into the generated noise over the specified GPS time interval, creating the mock data for further study.

%%%%%%%%%%%%%%%%%%%%%%%%%%%%%%%%%%%%%%%%%%%%%%%%%%
\subsection{Using \texttt{gwforge\_workflow}}
%%%%%%%%%%%%%%%%%%%%%%%%%%%%%%%%%%%%%%%%%%%%%%%%%%

Since the data generation process can be computationally expensive and time-consuming, users are encouraged to take advantage of \texttt{gwforge\_workflow}, which leverages the high-throughput computing framework HTCondor. This significantly accelerates the data generation process by abstracting all other housekeeping logic away from the user. Once the mock catalogues are generated and the different configuration files are created, the following commands can be executed:
\begin{widetext}
\begin{lstlisting}[style=bash]
# Activate Conda environment
conda activate gwforge-test || { echo "Failed to activate Conda environment." >&2; exit 1; }

# Set output directory
output_directory=forged-data

# Workflow submission script
gwforge_workflow \
    --gps-start-time 1895616018 \
    --gps-end-time  1895702418 \
    --output-directory "${output_directory}/output" \
    --noise-configuration-file "${output_directory}/xg.ini" \
    --bbh-configuration-file "${output_directory}/bbh-inject.ini" \
    --bns-configuration-file "${output_directory}/bns.ini" \
    --nsbh-configuration-file "${output_directory}/nsbh.ini" \    
    --workflow-name work \
    --submit-now
\end{lstlisting}
\end{widetext}

This will create an HTCondor meta-scheduler to manage dependencies between jobs and systematically parallelize the tasks, accelerating the mock data generation process. Fig~\ref{fig:data} shows 1024s of simulated data when assuming Gaussian noise, including signals from Population-A (See Sec.~\ref{sec:foreground} for details). As can be seen, the data will never be ``signal-free''.

% \RG{since you have a ``conda activate gwforge-test'' : lalsuite comes with a pre-determined environment.yml such that doing conda env create -f environment.yml automatically builds the needed environment. Does gwforge have something similar?}

% \KC{Yep. But as you suggested it will be better to just make a conda installable package such that all the requisite things get installed.}

%%%%%%%%%%%%%%%%%%%%%%%%%%%%%%%%%%%%%%%%%%%%%%%%%%
\section{Example usage of the mock data}\label{sec:application}
%%%%%%%%%%%%%%%%%%%%%%%%%%%%%%%%%%%%%%%%%%%%%%%%%%
%%%%%%%%%%%%%%%%%%%%%%%%%%%%%%%%%%%%%%%%%%%%%%%%%%
\subsection{Foreground Noise}\label{sec:foreground}
%%%%%%%%%%%%%%%%%%%%%%%%%%%%%%%%%%%%%%%%%%%%%%%%%%

\begin{figure}
    \centering
    \includegraphics[width=0.45\textwidth]{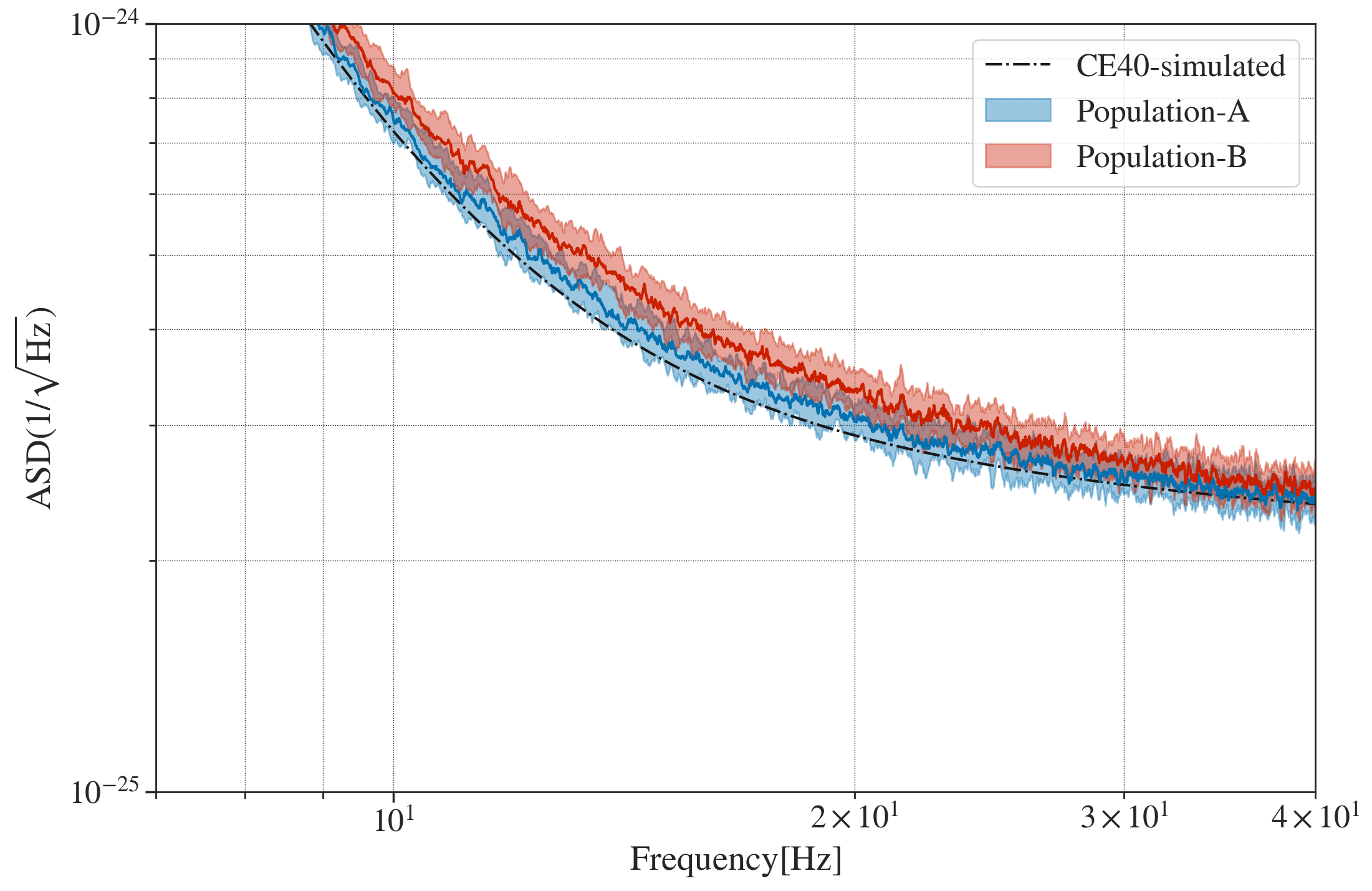}
    \caption{Comparison of Welch \ac{asd} estimates for different realizations of 512 seconds of \ac{ce}40 mock data. The black dotted line represents the ideal \ac{ce}40 design sensitivity curve. The red curve indicates the estimate when the noise includes signals from population A, while the blue curve represents the estimate when the noise includes signals from population B. Both the red and blue curves include a 90\% confidence band. The comparison demonstrates that depending on the number of signals, the \ac{asd} estimate can deviate from the expected noise \ac{asd} at 15Hz.}
    \label{fig:asd}
\end{figure}

\begin{figure}
    \centering
    \includegraphics[width=0.45\textwidth]{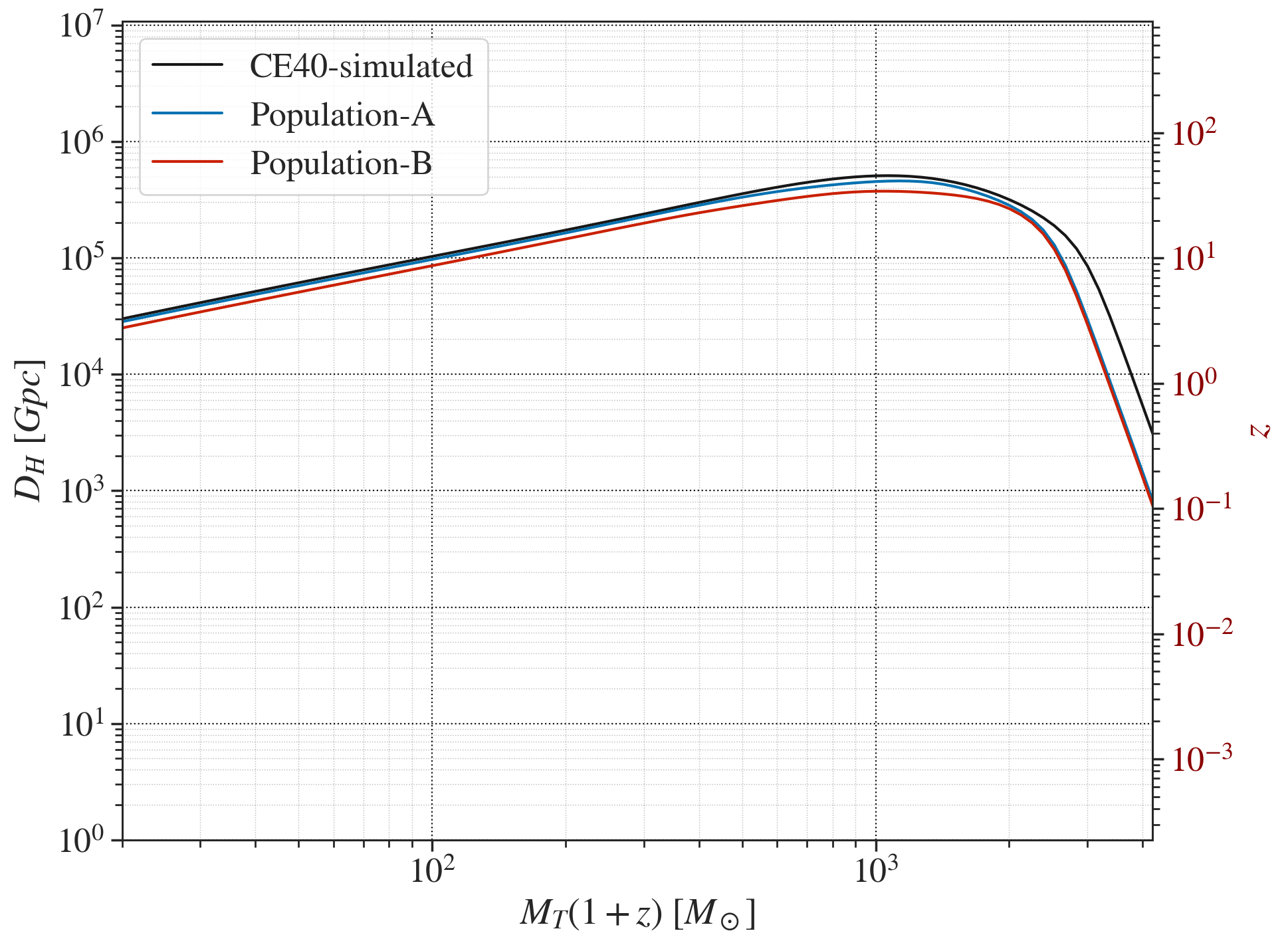}
    \caption{Comparison of horizon distance reach when using the median Welch \ac{asd} estimates. The black dotted line represents the ideal \ac{ce}40 design sensitivity curve. The red curve indicates the estimate when the noise includes signals from population A, while the blue curve represents the estimate when the noise includes signals from population B. The comparison shows that depending on the number of signals, the horizon distance of a \ac{xg} detector can be affected, especially for \acp{bbh} with a high detector frame (redshifted) total mass.}
    \label{fig:horizon}
\end{figure}

The increased bandwidth of \ac{xg} detectors will allow low-mass \ac{cbc} signals to remain within the detector's sensitive bandwidth for hours. Consequently, multiple signals will simultaneously appear in the data stream, the number of which will depend on the sources' astrophysical merger rate density. Characterised by their rapid frequency evolution, these signals will predominantly overlap at low frequencies and ``separate'' as they approach merger. Therefore, current modelled detection techniques can identify and measure their merger time to $\mathcal{O}(10ms)$ accuracy~\citep{Relton:2022whr}. As a result, unlike in LISA, where overlapping signals lead to source confusion, these signals will be resolvable and only contribute to \textit{foreground noise}.

However, as the number of overlapping signals per time-frequency bin is small, their combined contribution won't obey the central limit theorem. Therefore, the data cannot be assumed to follow a wide-sense stationary Gaussian process --- an assumption critical to the currently used noise \ac{psd} estimation technique such as the Welch averaging~\citep{Welch:1967oth, Allen:2005fk} and the BayesLine method~\citep{Littenberg:2014oda, Cornish:2014kda, Cornish:2020dwh}. As a result, using these techniques without modification can result in biased \ac{psd} estimates. %~\footnote{The Welch method presumes the data to be wide-sense stationary and estimates the noise \ac{psd} by dividing the data into equal-length overlapping segments, computing a modified periodogram for each segment, and averaging them. Conversely, BayesLine models the noise \ac{psd} using a linear combination of cubic splines and Lorentzians such that the residual whitened data follows a $\mathcal{N}(0,1)$ distribution.}.
Such a deviation can introduce systematic biases in parameter inference and reduce \ac{cbc} detectability.

To demonstrate this, we simulate multiple realisations of 512s mock \ac{ce}40 data using the \gwforge package and estimate the noise \ac{psd} using the Welch method. Mathematically, this method assumes the data to be wide-sense stationary and divides the discrete time domain data \( d[j] \) into \( K \) overlapping segments \( d_k[j] \) of length \( L \), each windowed by \( w[j] \) which for our purpose, is the Hann truncation function. The discrete Fourier transform \( \tilde{d}_k[i] \) of each windowed segment is computed using an FFT routine, and the periodogram \( P_k[i] \) is obtained as
\[
P_k[i] = \frac{1}{U} |\tilde{d}_k[i]|^2,
\]
where \( U = \sum_{n=0}^{L-1} w[j]^2 \). We, following~\citet{Allen:2005fk}, use mean-median averaging of the periodograms over all segments to obtain the final \ac{psd} estimate. This approach significantly reduces the variance of the spectral estimate due to glitches, thereby providing a more accurate representation of the data's power distribution across frequencies.

We assume two different source populations --- population-A quantified by \ac{bns} and \ac{bhns} merger rate density of $\mathcal{R}_\mathrm{BNS}^\mathrm{A}=320/\mathrm{Gpc}^3/\mathrm{yr},~ \mathcal{R}_\mathrm{BHNS}^\mathrm{A}=90/\mathrm{Gpc}^3/\mathrm{yr}$ respectively and population-B with merger-rate density of $\mathcal{R}_\mathrm{BNS}^\mathrm{B}=1700/\mathrm{Gpc}^3/\mathrm{yr},~ \mathcal{R}_\mathrm{BHNS}^\mathrm{B}=200/\mathrm{Gpc}^3/\mathrm{yr}$~\citep{LVK:2021duu, LVK:2024elc}. While overlapping \ac{bbh} signals will lead to larger foreground noise, their lower local merger rate and shorter duration in the detector band suggest that they will rarely contribute to the foreground noise. We therefore use $\mathcal{R}_\mathrm{BBH}=30/\mathrm{Gpc}^3/\mathrm{yr}$ in both the cases.

Fig.~\ref{fig:asd} shows the simulated and the estimated \ac{ce}40 \ac{asd}, with the latter obtained using the Welch method. It is estimated by dividing multiple-realisation of 512s of data into 16s subsegments and 50\% overlap, akin to that used in a \textsc{PyCBC} search. Consistent with previous works~\citep{Wu:2022pyg, Evans:2021gyd}, we find that the deviation is mainly concentrated between 6-30Hz, reaching $\gtrsim 10\%$ at 15Hz for the upper merger rate. This can reduce the detectability of \acp{bbh} with a high detector frame (redshifted) total mass, as shown in Fig.~\ref{fig:horizon}. In particular, depending on the binary's detector frame mass, the horizon distance~\footnote{The horizon distance, $D_H$, for our case, is the farthest luminosity distance to which we can confidently detect an optimally oriented, located, non-spinning symmetric mass \ac{bbh} assuming an optimal \ac{snr} threshold of 8.} loss to systems with $M_T(1+z) \gtrsim 1000M_\odot$ can be $\gtrsim 15\%$. Similar to the Welch estimate, the median BayesLine output, which is widely used in \ac{gw} \ac{pe}, will also show a deviation, unless a joint spectral and signal estimate is performed with precision (see Appendix~\ref{appx:bayes} for a discussion).

%%%%%%%%%%%%%%%%%%%%%%%%%%%%%%%%%%%%%%%%%%%%%%%%%%
    \subsection{Bright-siren cosmology}
%%%%%%%%%%%%%%%%%%%%%%%%%%%%%%%%%%%%%%%%%%%%%%%%%%

\begin{table}[htb]
    \caption{True value for the parameters used for the simulated \ac{imbh} binary. The values are inspired by the maximum likelihood values of GW190521 as inferred by~\citet{Islam:2023zzj}.}
    \centering
\scriptsize
\begin{tabular}{ l c c}
\hline \hline Parameter & Symbol & Value \\
\hline Mass ratio & $q$ & 0.95 \\
Detector-frame total mass & $M_T(1+z)$ & $273.83 M_{\odot}$ \\
Primary spin magnitude & $\chi_1$ & 0.83 \\
Secondary spin magnitude & $\chi_2$ & 0.96 \\
Primary tilt ${ }^{\mathrm{a}}$ & $\theta_1$ & 1.59 \\
Secondary tilt & $\theta_2$ & 1.96 \\
Azimuthal inter-spin angle & $\phi_{12}$ & 0.23 \\
Azimuthal precession cone angle & $\phi_{J L}$ & 0.36 \\
Effective aligned spin & $\chi_{\text {eff }}$ & -0.19 \\
Effective precessing spin & $\chi_p$ & 0.83 \\
Coalescence phase & $\phi$ & 1.06 \\
Polarization angle & $\psi$ & 1.28 \\
Coalescence GPS time & $t_c$ & 1242442967.43s \\
Right ascension & $\alpha$ & 4.39 \\
Declination & $\delta$ & 0.85 \\
Redshift & $z$ & 0.76 \\
Inclination angle & $\theta_{J N}$ & 2.09 \\
Hubble constant & $H_0$ &  $67.7 \mathrm{km}/\mathrm{s}/\mathrm{Mpc}$ \\
Matter density of Universe & $\Omega_m$ & $0.31$ \\
\hline \hline
\end{tabular}    
    \label{tab:true-values}
\end{table}

\ac{gw} signals from \ac{cbc} sources can be used as probes for cosmic expansion~\citep{Schutz:1986gp}. The luminosity distance, $D_L$ to a \ac{cbc} can be directly estimated using the \ac{gw} waveform $s(t\mid \boldsymbol{\theta}) \propto 1/D_L$ without needing external distance calibrators~\citep{Cutler:1994ys}. If the \ac{cbc}'s cosmological redshift, $z$, can be estimated by other means, then one can infer the local expansion rate, $H_0$ and other cosmological parameters~\citep{Holz:2005df, Nissanke:2009kt}, such as the fractional energy density of dark energy, $\Omega_\Lambda$ and dark matter, $\Omega_m$. For example, observing the \ac{em} counterparts from GW170817 allowed us to identify the source's host galaxy, NGC 4993, enabling the first direct measurement of $H_0$ using \acp{gw}~\citep{LIGOScientific:2017adf}.

Another promising \ac{em}-bright \ac{cbc} source is \acp{bbh} mergers occurring in gas-rich environments such as \ac{agn} disks~\citep{McKernan:2014oxa, Bartos:2016dgn, McKernan:2019hqs, Graham:2020gwr}. \acp{bbh} in such environments can efficiently merge due to gas torques and dynamical encounters, potentially producing detectable \ac{em} radiation during or after the merger. Such detection is feasible even against bright \ac{agn} disks, provided the disks are thin and relatively low in luminosity or if the merger remnant is ejected from the optically thick midplane and undergoes super-Eddington accretion.

\ac{xg} detectors promise to play a pivotal role in inferring cosmological parameters~\citep{Chen:2024gdn, Corsi:2024vvr}. To demonstrate how the mock data can be used to perform bright-siren cosmology, we analyse 8s of \ac{xg} detector data segment containing a GW190521-like binary that coincides with an \ac{em} counterpart with $\{\alpha_\mathrm{em}, \delta_\mathrm{em}, \mathrm{z}_\mathrm{em}\}$ same as the simulated signal. The data includes Gaussian noise and Population-A binaries, as introduced in the previous subsection. The binary and cosmological parameters used for the analysis are tabulated in Table~\ref{tab:true-values}. We use the Bayesian \ac{pe} library \bilby~\citep{Ashton:2018jfp}, waveform model \nrsur~\citep{Varma:2019csw} and off-the-shelf dynamic nested sampler \dynesty~\citep{Speagle:2019ivv}. We use the same waveform model for injection and recovery to prevent waveform systematics. Consistent with the mock data generation set-up, we use a minimum-frequency cutoff of $f_\mathrm{min}=6$Hz and perform a flare-unrestricted \ac{pe} with standard priors~\citep{LIGOScientific:2020iuh} and \ac{psd} calculated using the Welch method. 
% \RG{And what about the PSD? Also, maybe note down the SNR of the source, for cntext}

\begin{figure}[htb]
    \centering
    \includegraphics[width=0.45\textwidth]{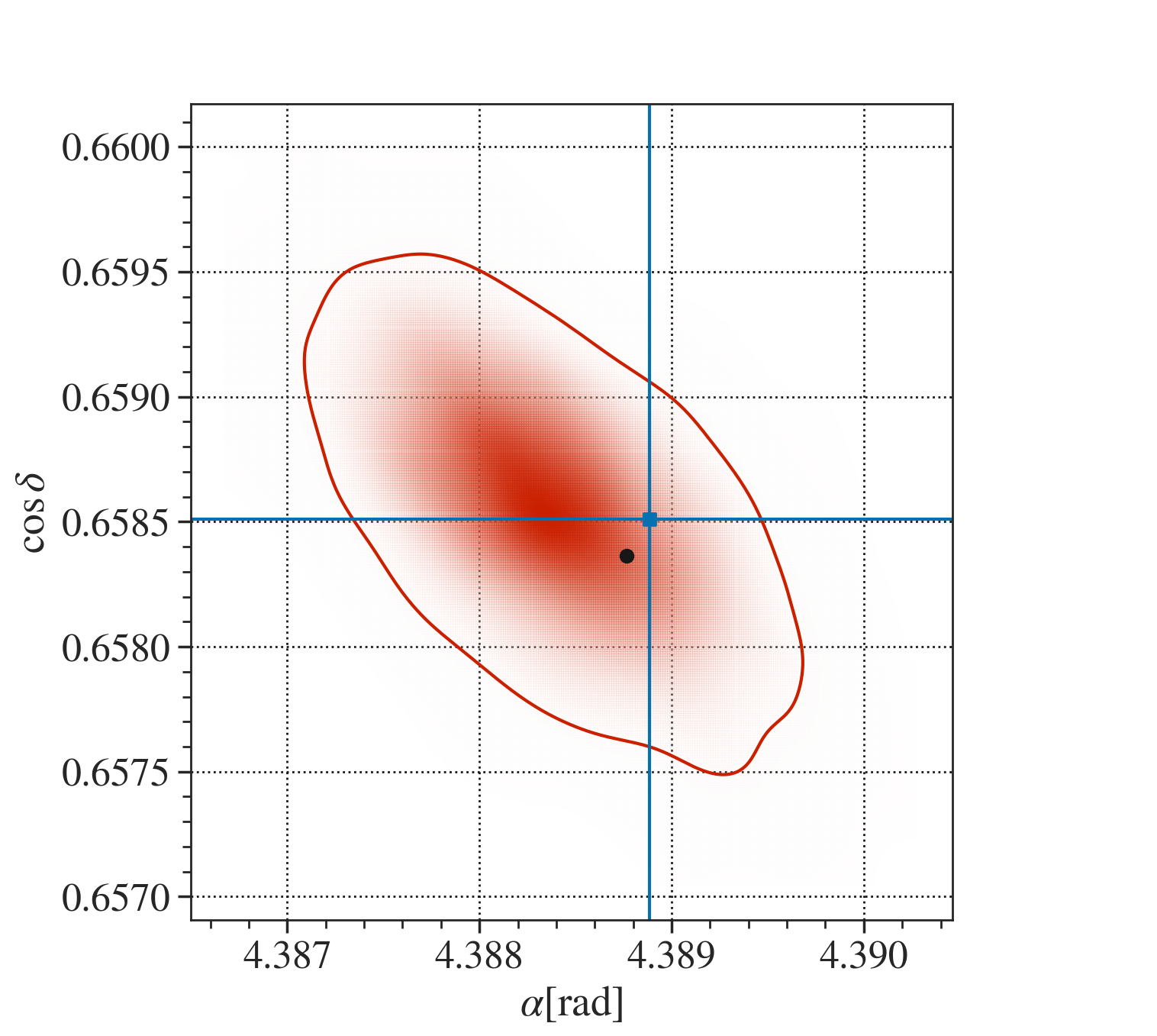}
    \caption{Posterior probability density for the sky-location of the GW190521-like signal in \ac{xg} network data. The blue square shows the sky location of the simulated binary, and the black dot shows the maximum likelihood sky location}
    \label{fig:gw190521-sky-coincidence}
\end{figure}

Fig.~\ref{fig:gw190521-sky-coincidence} shows the posterior probability distributions for the sky-location of simulated binary. As can be seen, the sky location of the flare and the \ac{gw} signal are consistent. In fact, the odds of a common source hypothesis $C$, against a random coincidence, $\mathcal{O}_{C/R} > 2 \times 10^7$, if we assume the prior odds, $\pi_{C/R}$,  to be unity.

\begin{figure}[h]
    \centering
    \includegraphics[width=0.45\textwidth]{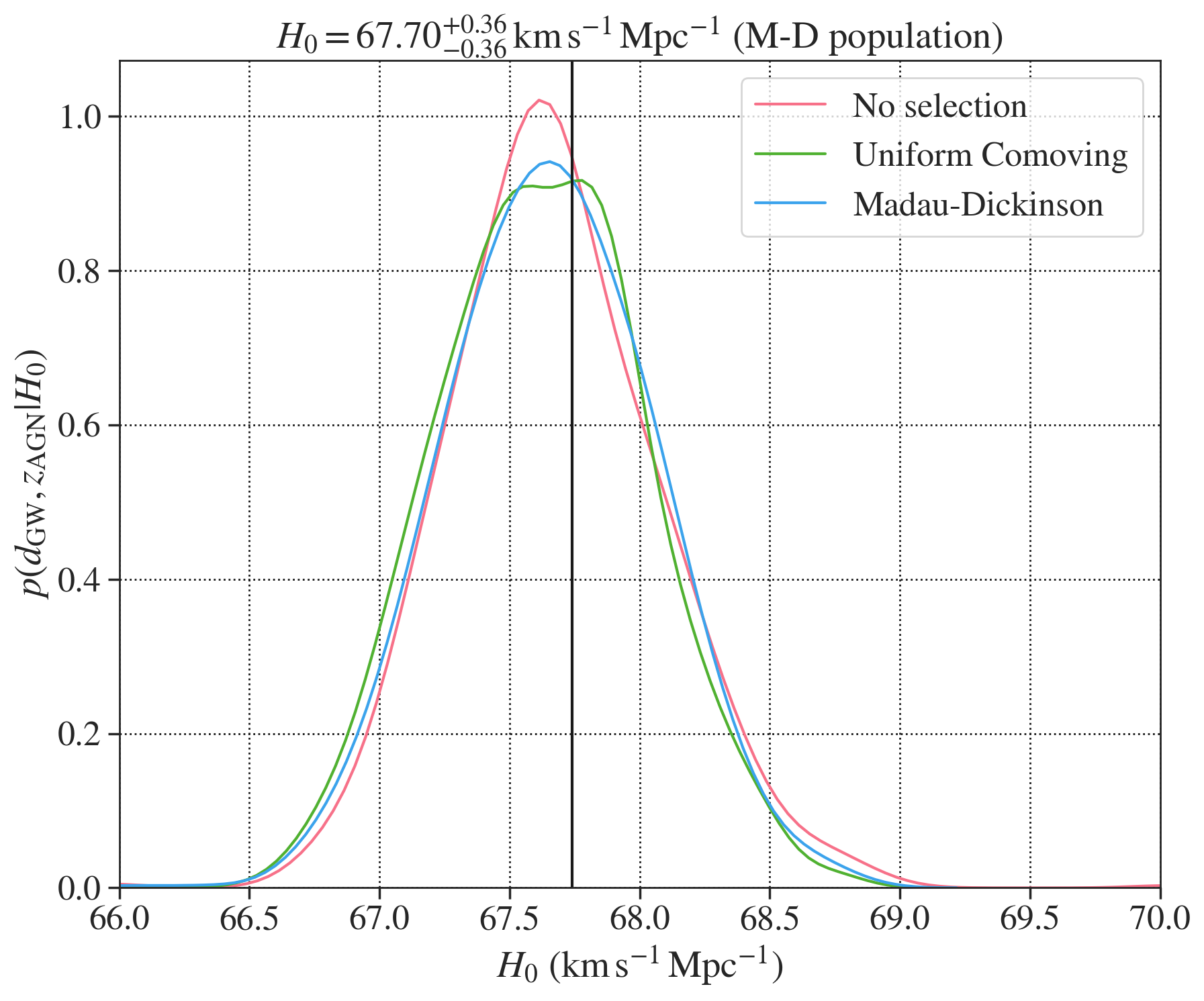}
    \caption{The posterior pdf of $H_0$ for the event under the assumption of flat $\Lambda$CDM cosmology and physical matter constraints from~\citet{Planck:2018vyg}. The red curve assumes no selection effect, while the green and blue curve assumes that the underlying \ac{bbh} population follows a Uniform in Comoving volume and star formation rate as modelled by~\citet{Madau:2014bja} respectively.}
    \label{fig:hubble}
\end{figure}

Having confirmed such an association, we use the obtained posterior samples to infer $H_0$~\citep{LIGOScientific:2017adf, Chen:2020gek} given the \ac{gw} and \ac{em} datasets, $d_\mathrm{gw}$ and $d_\mathrm{em}$:
\begin{equation}\label{eq:hubble}
\begin{aligned}
    p(H_0  &\mid d_\mathrm{em}, d_\mathrm{gw}) \propto  \, \pi(H_0) \\
    & \times \mathcal{L} \left( d_\mathrm{gw} \mid \alpha_\mathrm{em}, \delta_\mathrm{em}, D_L (z_\mathrm{em} \mid H_0) \right) \\
    & \times \frac{\pi_\mathrm{pop}(z_\mathrm{em} \mid H_0)}{\beta(H_0)}~.
\end{aligned}
\end{equation}
Here $\pi(H_0)$ is the prior on $H_0$, $\mathcal{L}$ is the marginalised \ac{gw} likelihood function evaluated at the distance and sky location of the electromagnetic source, $\pi_\mathrm{pop}(z_\mathrm{em} \mid H_0)$ is the assumed redshift distribution of the binary population and $\beta(H_0)$ is the fraction of binary population that will be simultaneously detected in \ac{gw} and \ac{em} window. 

\begin{figure}[htb]
    \centering
    \includegraphics[width=0.45\textwidth]{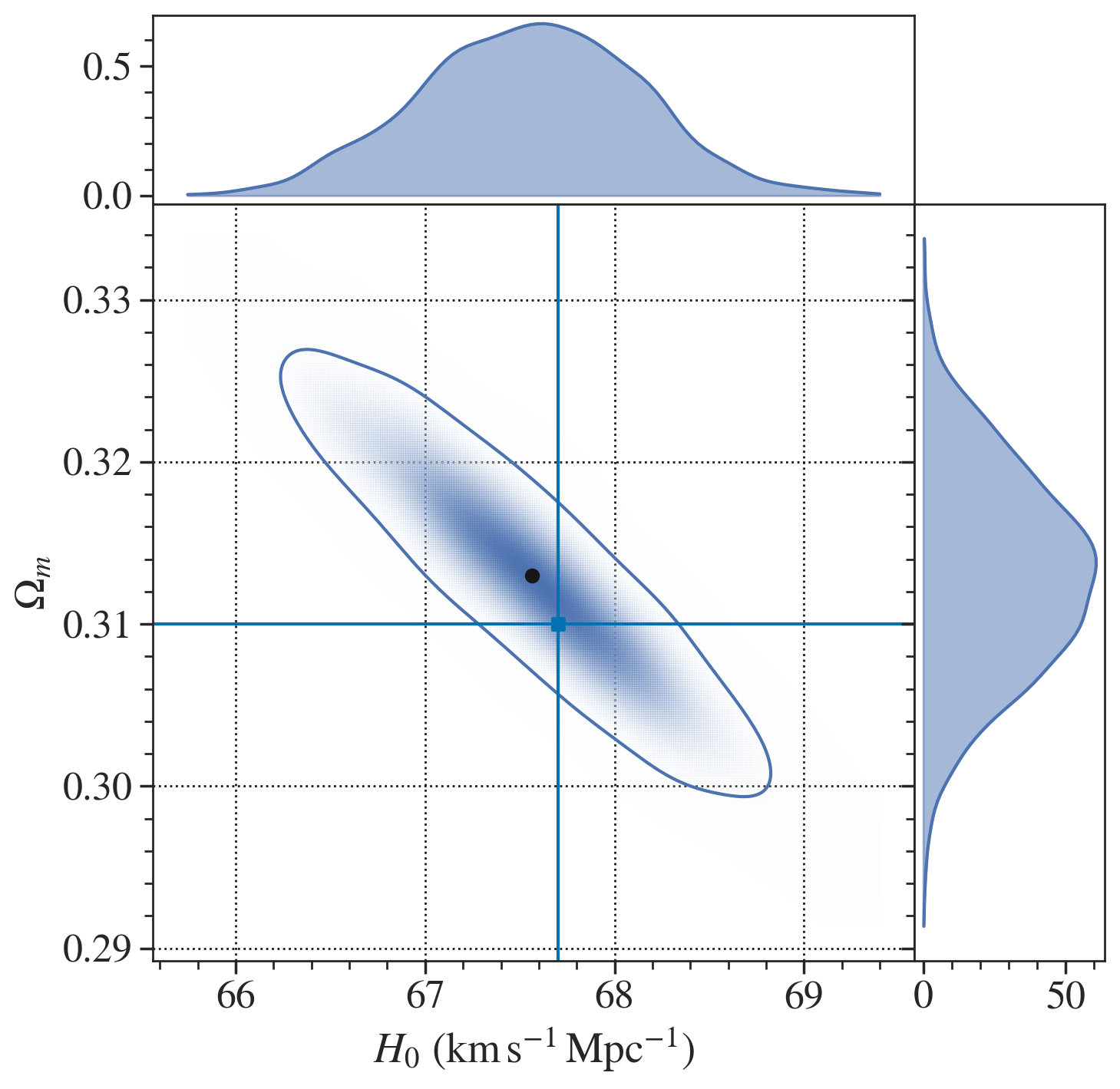}
    \caption{The joint posterior pdf of $H_0$ and $\Omega_m$ for the event using uniform priors for all the parameters in a flat $\Lambda$CDM $(w=-1)$ cosmology. The contour encloses 90\% CI. The blue square shows the cosmological parameters used for the simulations, while the black dot shows the maximum likelihood sky location.}
    \label{fig:joint-hubble}
\end{figure}

Fig~\ref{fig:hubble} shows our results, assuming a flat $\Lambda$CDM cosmology and matter constraints from \citet{Planck:2018vyg}. We assume $\pi(H_0)=\mathcal{U}(35, 140)\mathrm{km/s/Mpc}$ and consider two redshift distribution models: one that assumes that \ac{bbh} merger rate is constant up to redshift $z=10$ and the other that follows the Madau-Dickinson star-formation rate used to simulate the binary population. For both cases, we first estimate $\beta(H_0)$ following~\citet{farr2018} and then infer $H_0$ to be $67.7^{+0.36}_{-0.36}\mathrm{km/s/Mpc}$ at 68\% CI, thereby being consistent with the value used in simulating the binary population. It shows that if the uncertainty in luminosity distance and the counterpart redshift measurement is small, \ac{xg} \ac{em}-bright sources can precisely measure $H_0$, irrespective of the underlying population distribution assumption.

In Fig~\ref{fig:joint-hubble}, we present the measurements of $\Omega_m$ and $H_0$, when using flat priors in the range $H_0=[35,140]\mathrm{km/s/Mpc}$, $\Omega_m=[0,1]$ and assuming $\Lambda$CDM cosmology. For this, we obtain the joint posterior on $(H_0, \Omega_m)$ by modifying Eq.~\eqref{eq:hubble} to the following:
\begin{equation}
\begin{aligned}
    p(H_0, \Omega_m  &\mid d_\mathrm{em}, d_\mathrm{gw}) \propto  \, \pi(H_0, \Omega_m) \\
    & \times \mathcal{L} \left( d_\mathrm{gw} \mid \alpha_\mathrm{em}, \delta_\mathrm{em}, D_L (z_\mathrm{em} \mid H_0, \Omega_m) \right) \\
    & \times \frac{\pi_\mathrm{pop}(z_\mathrm{em} \mid H_0, \Omega_m)}{\beta(H_0, \Omega_m)}~.
\end{aligned}
\end{equation}
Consistent with Fig.~\ref{fig:hubble}, the maximum likelihood point is close to the true value. However, as expected, the posterior for $H_0$ is broader.

%%%%%%%%%%%%%%%%%%%%%%%%%%%%%%%%%%%%%%%%%%%%%%%%%%
    \subsection{PE biases due to waveform systematics}
%%%%%%%%%%%%%%%%%%%%%%%%%%%%%%%%%%%%%%%%%%%%%%%%%%
\begin{figure}[htb]
    \centering
    \includegraphics[width=0.49\textwidth]{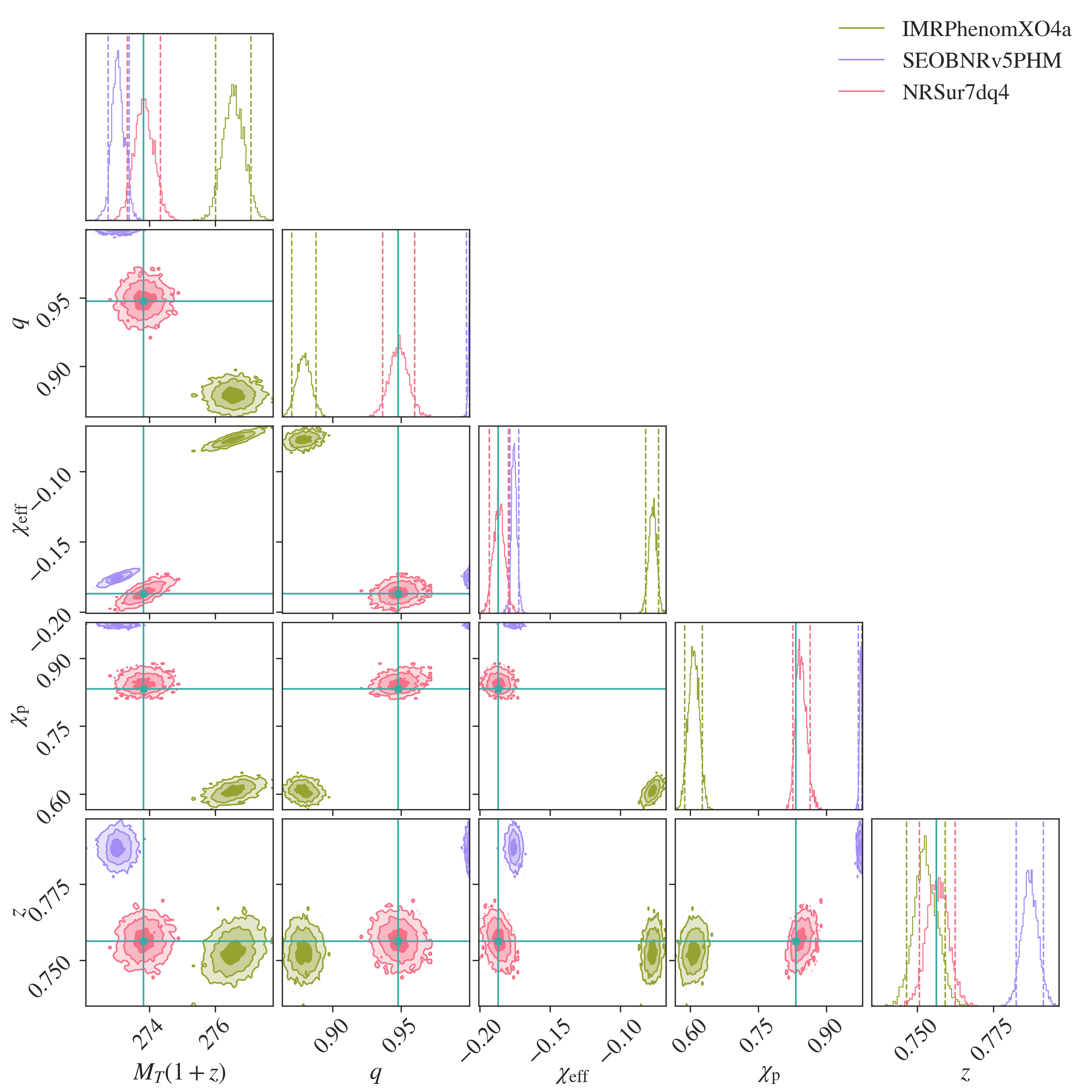}
    \caption{Comparison corner plot for the posteriors obtained for the redshifted total mass, mass ratio, effective aligned and precessing parameters and the redshift for the GW190521-like signal when using three different waveform approximants. The cyan lines correspond to the true values.}
    \label{fig:waveform-systematics}
\end{figure}

\acp{gw} from \acp{cbc} carry characteristic information about the astrophysical properties of compact objects, such as their masses and spins. A significant challenge in inferring these parameters is the systematic error introduced by the waveform approximants. If two waveform models, $\mathcal{W}_1$ and $\mathcal{W}_2$ do not agree for a particular choice of $\boldsymbol{\theta}$, i.e. $\delta h = h_1(t \mid \boldsymbol{\theta}, \mathcal{W}_1) - h_2(t \mid \boldsymbol{\theta}, \mathcal{W}_2) \not\rightarrow 0$, then it can significantly skew \ac{pe} outcomes. This is especially problematic if the difference exceeds the detector noise level.

More rigorously, if the two waveform models are not \textit{faithful} up to a particular level determined by the signal's \ac{snr}, then the  disagreement between the waveform models will lead to differences in \ac{pe} outputs.  This issue is particularly critical in \ac{xg} era, where we anticipate observing over $10^5$ compact binary signals per year, few of which can have a \ac{snr} greater than 100. Since the statistical uncertainty of parameter estimates decreases with an increase in \ac{snr}, the measurability of signal parameters can be significantly affected if the current waveform models are used, particularly when analysing high \ac{snr} events~\citep{Purrer:2019jcp, Dhani:2024jja, Kapil:2024zdn}. Consequently, it is crucial to understand how waveform systematics can lead to incongruous results.

While a detailed study is deferred to future work, we provide an illustrative example of using the simulated mock data to identify biases in \ac{pe} due to waveform systematics. We note that these studies must be performed in zero-noise so that no systematics are introduced due to noise. Here, we give an example where the noise is Gaussian and includes population-A signals.

To this end, we re-analyse the mock data segment containing the GW190521-like \nrsur signal with parameters tabulated in Table~\ref{tab:true-values} with three different state-of-the-art waveform models, namely \nrsur, \texttt{IMRPhenomXO4a} and \texttt{SEOBNRv5PHM}~\citep{Ramos-Buades:2023ehm}. For each of these analyses, we use \bilby, employ standard priors and the standard likelihood function for \ac{gw} transients~\citep{Finn:1992wt} and evaluate it from a fixed minimum frequency of $f_\mathrm{min}=6$Hz identical to the reference frequency choice.  

Fig.~\ref{fig:waveform-systematics} shows a corner plot comparing a few binary parameters. As can be seen, depending on the waveform approximant, the recovered parameters can be significantly different from the injected value. This is because these waveform models differ in their construction; therefore, measurements relying on them can have systematic uncertainties. This has not been an issue for almost all second-generation \ac{gw} candidates, as differences caused by different approximant choices have been lower than those induced by noise fluctuations. But for loud \ac{xg} events, this can introduce significant biases as the statistical uncertainty of parameter estimates decreases with an increase in \ac{snr}~\citep{Kapil:2024zdn, Dhani:2024jja}. Standard approaches to account for modelling errors, such as combining posterior for analysis performed using different approximants to yield a joint posterior that is effectively marginalised over the uncertainty inherent in the waveform models, won't alone be able to account for the errors and biases that occur during the recovery of source parameters due to waveform modelling choices.

%%%%%%%%%%%%%%%%%%%%%%%%%%%%%%%%%%%%%%%%%%%%%%%%%%
\subsection{Other possible challenges}
%%%%%%%%%%%%%%%%%%%%%%%%%%%%%%%%%%%%%%%%%%%%%%%%%%

The data generated using \gwforge can be used to address several other challenges, some of which are summarised below:
\begin{enumerate}
    \item \textbf{Testing faster and more robust search and \ac{pe} algorithms:} The \ac{xg} detectors are expected to observe hundreds of thousands of binary signals annually. Developing optimal signal identification methods and rapid parameter estimation algorithms tailored to compact binary analysis will be essential for extracting science from the signal-rich dataset. In this context, the traditional requirement of achieving a $\gtrsim 5\sigma$ confidence level may no longer be necessary. Instead, search design requirements can be adapted to enable faster detection and \ac{pe} algorithms can be developed and/or tuned for faster and more precise estimation of signal properties.

    Moreover, non-stationary noise due to glitches and foreground noise will necessitate a joint analysis of the signal of interest and the remaining data. In fact, current search techniques can be replaced by coherent Bayesian search algorithms that will allow for the identification of potential candidates, give rapid, fully coherent parameter estimates of the signal source, and assign astrophysical probabilities of the candidate events~\citep{Isi:2018vst}. The simulated datasets will allow the \ac{gw} community to develop, test, and refine algorithms to meet these scientific objectives.

    \item \textbf{Measurement of \ac{ns} \ac{eos}:} \acp{gw} from binaries consisting of a pair of \ac{ns}, or a \ac{ns} and a \ac{bh} can be used to measure the tidal deformability of \acp{ns} and constrain the nature of matter at supernuclear densities~\citep{Hinderer:2009ca, Damour:2012yf, DelPozzo:2013ala}. Given that these effects appear in the fifth \acl{pn} order, this will be best measured for signals with \ac{snr} of order thousands, akin to those expected to be observed in \ac{xg} era. The \gwforge dataset will contain such signals depending on the initial conditions and can be used to verify how precisely and accurately one can measure the \ac{ns} \ac{eos}. Further, it can be used to quantify any observed bias due to the foreground noise. Also, depending on the waveform model used, the post-inspiral portion of the signals may also be present in the simulated dataset which can be used to study how well we can constrain the presence of phase transitions at several times the nuclear densities~\citep{Radice:2016rys, Most:2018eaw, Most:2019onn, Prakash:2023afe}.        

    \item \textbf{Testing rapid pre-merger sky localisation algorithms:} Low latency detection and pre-merger localisation of potential \ac{bns} signals, especially those within the reach of electromagnetic observatories, will be an important scientific goal for \ac{xg} \ac{cbc} algorithms as they can allow for a multiband understanding of \acp{bns}. The \gwforge dataset can be used to mimic online analysis. It will allow the community to combine multi-messenger observation of future mergers to more precisely determine the \ac{ns} \ac{eos} and learn about the engine powering gamma-ray bursts, kilonovae and the formation of heavy elements.     
    
    \item \textbf{Estimate the strength of the stochastic background due to compact binary population:} The aggregate of all sources, particularly binary neutron stars will contribute to foreground noise which could interfere with the study or the establishment of upper limits on the cosmological \ac{gw} background. Therefore, it is crucial to either remove or account for this foreground to measure the properties of a primordial background accurately. \gwforge datasets will facilitate extensive testing of \ac{xg} algorithms on large and realistic samples.

    \item \textbf{Investigating Deviations from \ac{gr} and False Violations:} \gwforge allows injecting non-GR signals through a \pycbc plugin, enabling users to introduce and analyze these signals. This feature facilitates studies on the efficacy of conventional tests in detecting deviations from GR. Additionally, users can inject GR signals to investigate scenarios where tests of GR may fail. Such failures could arise due to waveform systematics, sub-optimal detector noise modelling, and various astrophysical aspects. By examining both non-GR and GR signals, \gwforge dataset will allow us to perform mock tests of GR in \ac{xg} era.

    \item \textbf{Determining Merger Rates and Reconstructing the Underlying Distribution of Source Populations:} Given the anticipated high number of detections, utilizing current hierarchical Bayesian inference techniques to determine the population properties of a \ac{cbc} type may become impractical. However, the sheer volume of signals allows for an alternative approach: multiplying the posteriors of individual events of a particular source type to infer population properties. The \gwforge dataset will enable the validation of this hypothesis and facilitate the testing of faster and more accurate Bayesian inference techniques.

\end{enumerate}

\section{Conclusion}\label{sec:conclusion}
%%%%%%%%%%%%%%%%%%%%%%%%%%%%%%%%%%%%%%%%%%%%%%%%%%

In the upcoming decades, \ac{xg} detectors are expected to replace the current \ac{gw} detection facilities. With longer baselines and new infrastructures, these detectors can make transformative discoveries across astronomy, physics and cosmology. However, the ``richer'' \ac{gw} detector data will pose new computational, physical and astrophysical challenges, which may not be straightforward to mitigate. 

For example, the sheer volume of signals will require developing novel signal identification and extraction algorithms. Non-stationarity due to foreground noise will pose additional challenges, especially in noise modelling, complicating subsequent analysis.  Furthermore, detecting loud signals will necessitate the development of improved waveform models to achieve unprecedented precision in parameter estimations and avoid false indications of deviations from \ac{gr}. 

To simulate these challenges, we introduce \gwforge, a user-friendly and lightweight Python package designed to generate mock data for \ac{xg} detectors. Through various data simulation examples, we demonstrate the capabilities of \gwforge. These examples include examining foreground noise from overlapping signals with intersecting time-frequency tracks, demonstrating the potential of bright sirens for $H_0$ measurement, and discussing parameter estimation biases due to waveform systematics. We also highlight a few potential challenges that can be addressed using the simulated dataset. These demonstrations underscore the potential applications of \gwforge in supporting the development of new analysis methods and preparing the scientific community for the next era of gravitational wave astronomy.

% In future

%%%%%%%%%%%%%%%%%%%%%%%%%%%%%%%%%%%%%%%%%%%%%%%%%%
\section*{Acknowledgements}\label{sec:acknowledgements}
%%%%%%%%%%%%%%%%%%%%%%%%%%%%%%%%%%%%%%%%%%%%%%%%%%%
The author thanks B.S. Sathyaprakash, Ian Harry, Rossella Gamba and  Ish Gupta for their insightful comments and valuable suggestions. He extends special thanks to Sophie Hourihane for her assistance with BayesLine. Heartfelt appreciation is also given to Lorine Talhaoui, Tobias Forge, Papa Nihil, and N. Ghouls for their continuous inspiration throughout the development of this work. 

The author acknowledges the generous support of the National Science Foundation (NSF) through grant PHY-2207638. He is also grateful for the computational resources provided by the Gwave (PSU) cluster, which were essential for the numerical work conducted in this research.

The \texttt{gwforge} package extensively utilises the following \texttt{gw} packages: \bilby~\citep{Ashton:2018jfp}, \pycbc~\citep{Usman:2015kfa}, \texttt{gwpy}~\citep{gwpy}, and \texttt{Lalsuite}~\citep{lalsuite}. Furthermore, various \textsc{Python} packages such as \texttt{numpy}~\citep{harris2020array}, \texttt{matplotlib}~\citep{Hunter:2007}, \texttt{corner}~\citep{corner}, and \texttt{seaborn}~\citep{Waskom2021} are used in this work.

\appendix

%%%%%%%%%%%%%%%%%%%%%%%%%%%%%%%%%%%%%%%%%%%%%%%%%%
\section{BayesLine PSD estimate}\label{appx:bayes}
%%%%%%%%%%%%%%%%%%%%%%%%%%%%%%%%%%%%%%%%%%%%%%%%%%

\begin{figure}
    \centering
    \includegraphics[width=0.45\textwidth]{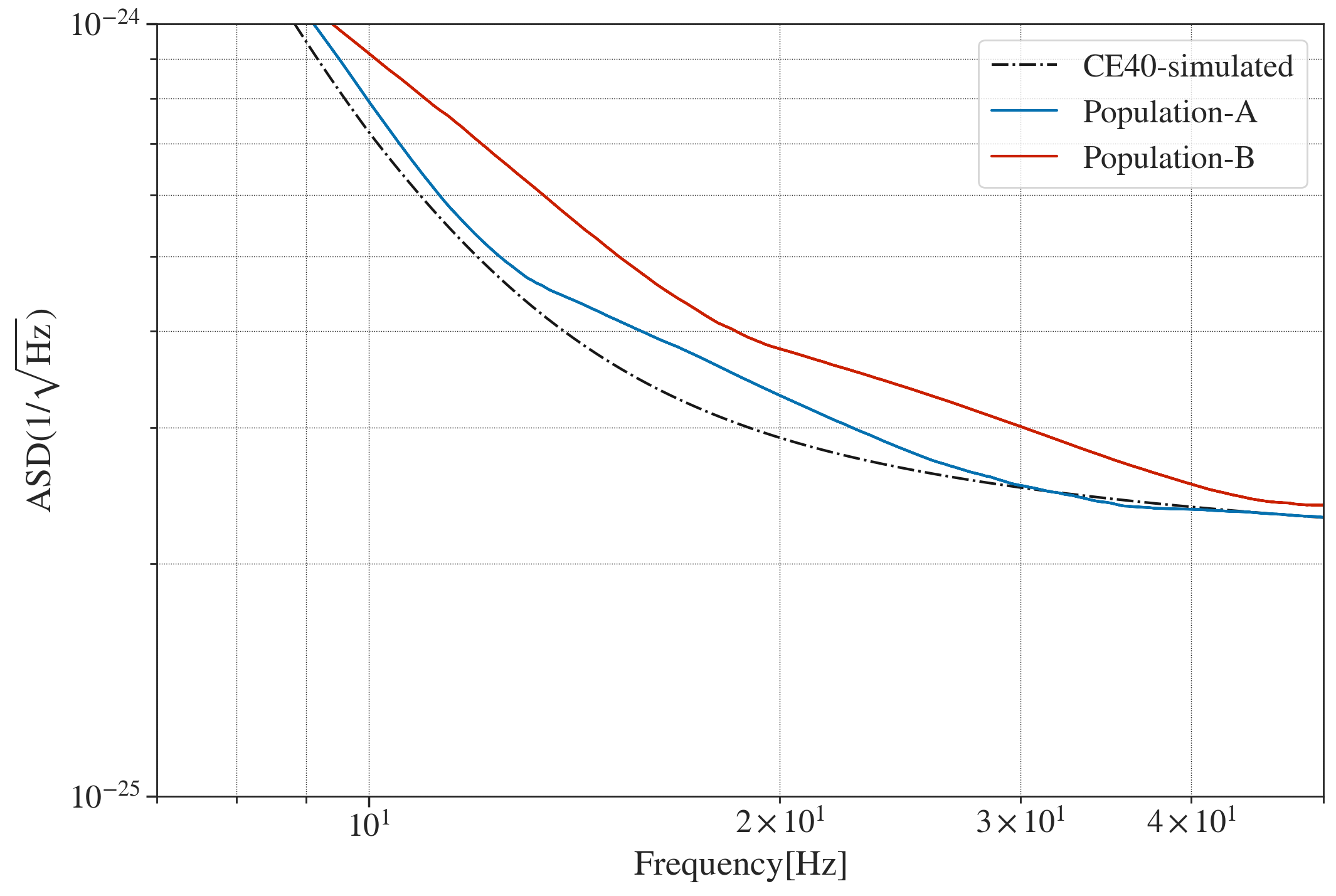}
    \caption{Comparison of BayesLine median \ac{asd} estimates of 512 seconds of \ac{ce}40 mock data. The black dotted line represents the ideal \ac{ce}40 design sensitivity curve. The red curve indicates the estimate when the noise includes signals from population A, while the blue curve represents the estimate when the noise includes signals from population B. The comparison demonstrates that depending on the number of signals, the \ac{asd} estimate can deviate significantly at lower frequencies from the expected noise.}
    \label{fig:bayesline-asd}
\end{figure}

The Welch method for estimating noise power spectral density (PSD) generates an ``off-source'' noise PSD by median-averaging the spectral estimates of neighbouring data segments. However, it is not widely used in \ac{gw} \ac{pe} due to its sensitivity to non-Gaussian transients or ``glitches''. Moreover, even for wide-sense stationary Gaussian noise, the data whitened by the Welch method doesn't produce a $\mathcal{N}(0,1)$ distribution, leading to biased \ac{pe} estimates~\citep{Talbot:2020auc, Chatziioannou:2019zvs}.

Hence, current \ac{lvk} analyses use the ``on-source'' Bayesian spectral estimates, such as those provided by the \textsc{BayesLine} algorithm~\citep{Littenberg:2014oda, Cornish:2014kda, Cornish:2020dwh}. This method approximates the noise \ac{psd} by generating a parameterised fit involving a combination of splines and Lorentzians. This has yielded unbiased and statistically consistent results, with whitened residuals that follow a $\mathcal{N}(0,1)$ distribution. 

Furthermore, this methodology, originally developed for joint spectral and signal inference, will be indispensable in the upcoming \ac{xg} era, where no signal-free off-source data will be available. However, current \ac{lvk} analysis uses sequential modelling, which may lead to notable discrepancies from joint analyses as detector sensitivities improve. This discrepancy becomes more pronounced, particularly for ground-based \ac{xg} detectors, as the sum of foreground noise won't follow a Gaussian distribution.

To illustrate that, we re-compute the noise \ac{psd} of a 512s segment of data as discussed in Sec.~\ref{sec:foreground} using the \textsc{BayesLine} algorithm. Fig.~\ref{fig:bayesline-asd} shows our result. Like the Welch estimate, the foreground noise significantly affects noise characterisation at lower frequencies, with a more pronounced impact observed for population-B data, which assumes upper merger rates for \ac{bns} and \ac{bhns} systems. This can affect \ac{pe} results, potentially broadening constraints on parameters like chirp mass ($\mathcal{M}_c$) and/or symmetric mass ratio ($\eta$) for longer signals while not significantly affecting the measurability of parameters appearing at higher \ac{pn} orders (see \citet{Johnson:2024foj} for a possible way to mitigate this). However, this may substantially impact binary parameter measurements for short-lived signals, such as \ac{imbh} binaries. A detailed investigation is deferred to future work.



\bibliography{PASPsample631}{}
\bibliographystyle{aasjournal}

\end{document}

% End of file `PASPsample631.tex'.